# A NOVEL APPROACH TO FULLY PRIVATE AND SECURE AUCTION: A SEALED-BID KNAPSACK AUCTION

**Maged Hamada Ibrahim**
Department of Electronics, Communications and Computers,
Faculty of Engineering, Helwan University,
1, Sherif st., Helwan, Cairo, Egypt.

**ABSTRACT**
In an electronic auction protocol, the main participants are the seller, a set of trusted auctioneer(s) and the set of bidders. In this paper we consider the situation where there is a seller and a set of *n* bidders intending to come to an agreement on the selling price of a certain good. Full private or bidder-resolved auction means that this agreement is reached without the help of trusted parties or auctioneers. Therefore, only the seller and the set of bidders are involved, the role of the auctioneers becomes obsolete in this case. We propose a new – simple and secure – technique for the design of a full private sealed-bid auction protocol. We employ the well known mathematical proposition, the knapsack problem which was used by Merkle and Hellman [1] in the design of their asymmetric public-key knapsack trapdoor cryptosystem. Up to our knowledge, the knapsack problem has not been considered before in the design of electronic auctions. We also employ an efficient (1-out-of-*k*) oblivious transfer of strings for secure data transfer between the seller and the bidders (e.g., [2]). At the end of the protocol, the seller knows the set of prices selected by the bidders, yet he doesn't know which bid belongs to which bidder until the winning bidder announces himself and proves his case by opening a secret code corresponding to the highest price. Our protocol is a 1[st] price and automatically a 2[nd] price auction as well, since the winning bidder can pay the 2[nd] highest price – indicated by a flag – according to the public auction predefined rules. We give the protocol for honest but curious participants then we show how to detect malicious behavior of the participants by employing a one way function with a suitable homomorphic property.

## 1. INTRODUCTION

Auctions become a major phenomenon in the field of electronic commerce. In an electronic auction, there is a seller that wants to sell a good; he defines a set of prices (several hundred tokens) for his good. There is a set of bidders who are willing to bid for this good; each bidder makes his selection for the price. At the end of the protocol, the bidder with the highest bid wins the auction and pays the highest price or sometimes the second highest price according to the auction public rule of payment. Some properties must be satisfied by any auction protocol. Fast execution: this is a common desired property in any computer network protocol. Economic design: the auction must be designed on solid economic principles, the bidders have to bid as they truly value the item based on their own valuation or indifference price. Privacy: no information is revealed about any bidder's bid during or after the auction is completed, only the winning bidder and the corresponding selling price is known at the end of the protocol. Anonymity: no information is revealed about any bidder's identity, the protocol is completely anonymous until the winning bidder announces himself to the public or to the seller with a proof that he is the intended winner, and no information about the identity of the other bidders is revealed. At the end of our protocol, the seller knows the set of prices selected by the bidders yet he doesn't know which price belongs to which bidder. Since no information is revealed about the identity of any bidder or his corresponding bid, privacy and anonymity are attained.

Auctions can be classified into three main categories, these are: increasing price (English auction), decreasing price (Dutch auction) and sealed-bid auction. In an English auction, the good is offered at increasing prices. Initially, the good is offered at $T$ tokens, in time slot *i*, it is offered at $(T+i\delta)$ tokens where $\delta$ is a function of several factors such as the previous bids. This type of auction has several disadvantages, the time required to conduct the auction is proportional to the price at which the item is sold, and the communication costs may grow super linearly in the ultimate price at which the item is sold. This type of auction leaks enormous amount of information, an observer can deduce information about the price that each party is willing to pay for this good and hence; true valuation, privacy and anonymity are lost. Dutch auction is similar to the English auction except the way the price varies over time. In this case, the price decreases, that is, at time slot *i*, the good is offered at $(T-i\delta)$ tokens. The first bidder who is willing to bid wins the auction. Hence, this type of auction provides maximum privacy and anonymity. However, as in the case of English auction, it is time consuming. In a sealed-bid auction, each party sends a sealed bid to a trusted auctioneer who opens all bids; the auctioneer sells the good to the bidder of the highest bid. The sealed-bid auction is very attractive, since, it can execute in one round of communication between the bidders and the auctioneers and hence it is very fast. However, the trusted auctioneers are its main disadvantage. Our protocol is a sealed-bid auction without the help of any auctioneers, that is, only the seller and the set of bidders are able to come to an agreement on the selling price.

This paper is organized as follows: In section 2, we give a brief description of the related work in the field and the contributions of this paper. In section 3, we give a bird's eye view on our protocol. Section 4 presents the basic tools used in the protocol. Section 5 presents a concrete description of the protocol in the case of honest but curious participants to clarify the basic steps; also, we include a simple numerical illustrative example. Section 6 presents the full protocol to detect any malicious behavior attempted by the participants.

## 2. RELATED WORK AND CONTRIBUTIONS

Franklin and Reiter [3] introduced the basic problem of sealed-bid auction but disregarded the privacy of bids after the auction is finished. Many secure auction protocols have been proposed, some of which are not suitable for the execution of a $2^{nd}$ price auction [3, 4, 5, 6, 7, 8, 9, 10, 35]. Another category of the work in this field is based on threshold cryptography and relies on the existence of a set of auctioneers where at most one third of them are not trusted [11, 12, 13, 14, 15]. Some of the proposed protocols rely on the existence of a third party that is not fully trusted. In [16, 17] the third party may not collude with any bidder, while in [18, 19] the third party and the auctioneers may collude. In [20, 21] there is a trusted third party that holds the key for a public-encryption algorithm. The recent work of Felix Brandt [22, 23, 24, 25] introduces a novel kind of secure and private auction where information is shared among bidders and only the seller and the set of bidders are incorporated in the protocol. In this paper we propose a protocol for electronic sealed-bid auction which is very simple with much lower complexity and satisfies full privacy and security. The protocol proposed in this paper satisfies the following properties:

- *Full privacy or bidder-resolved*: There are no trusted parties or auctioneers involved in the protocol, only the seller and the set of bidders are involved and are able to come to an agreement on the selling price.
- *Private-bids*: No information is revealed about any bidder's bid, at the end of the protocol the seller knows the set of prices chosen but he does not know which bid belongs to which bidder.
- *Anonymity*: No information is revealed about the identity of any of the bidders until the winning bidder of the highest bid identifies himself to the seller and he is able to prove his case by opening his secret code corresponding to the highest bid.
- *Fast execution*: Our protocol is a sealed-bid auction protocol; therefore, the protocol is fast compared to other protocols that depend on time slotted tokens such as English and Dutch auctions. Also, the novel attempt of employing the knapsack mathematical proposition speeds up the execution of the protocol over previous sealed-bid auction protocols.
- *Correctness*: the winning bidder and the selling price are determined correctly.
- *Applicability*: Our protocol applies to the $1^{st}$ price and $2^{nd}$ price auction as well.

## 3. THE OUTLINES OF OUR PROTOCOL

There is a seller that wants to sell a good; he selects a set $\mathcal{P}$ of $k$ prices (tokens) for this good where, $\mathcal{P} = \{p_1, ..., p_k\}$ and arranges these prices in an ascending order. The seller also selects a set $C$ of $k$ super-increasing secret values, $C = \{c_1, ..., c_k\}$, $c_i \in Z_q^*$ where $q$ is a large prime subject to the condition that $q > \sum_{i=1}^{k} c_i$. By the term 'super-increasing', we mean that $c_i > \sum_{j=1}^{i-1} c_j$ $\forall i = (1,...,k)$. The seller selects a vector of $n$ random and independent integers $\mathcal{R} = \{r_1, ..., r_n\}$ such that $\sum_{i=1}^{n} r_i = 0 (\bmod q)$, $r_i \in Z_q^*$, which will be used to randomize the distribution of $C$. Each price $p_i$ is assigned the secret code, $c_i$. There is also a $k$-bit vector $\mathcal{F} = \{f_1, ..., f_k\}$ where, $f_i \in \{0,1\} \forall i = (1,...,k)$, at the start of the protocol, $f_i = 0 \forall i = (1,...,k)$.

There is a set $\mathcal{B}$ of $n$ bidders, $\mathcal{B} = \{B_1, ..., B_n\}$. Each bidder $B_j \in \mathcal{B}$ secretly chooses a price $p_{i_j} \in \mathcal{P}$ that he is willing to pay for this good. The seller interacts with each bidder $B_j$ to secretly transfer the secret code $c'_{i_j} = c_{i_j} + r_j \bmod q$ corresponding to the bidder's selected price $p_{i_j}$. The seller must not know any information about the choice made by any of the bidders. Also, any bidder must not know any information about any secret code other than the one corresponding to his chosen price. Our solution to this problem is the employment of a secure $OT_k^1$ oblivious transfer of strings.

After each bidder possesses the secret code corresponding to his chosen price, the bidders – jointly and securely – compute the sum of the secret codes they have and delivers the result to the seller, we must satisfy that no bidder knows any information about any other bidder's bid (secret code). Also, the seller must not know which bid belongs to which bidder. This can be achieved through a trivial additive joint secret sharing scheme. Simply, each bidder splits his secret code into $n$ additive shares and privately sends a share to each other bidder. Each bidder then sums what he has and the result is delivered to the seller. The seller, by his role, sums what he receives from the bidders to compute the knapsack value and starts to solve the knapsack problem for the value he receives from the bidders to determine the set of flags, $\mathcal{F}$. As a result of the solution of the knapsack problem, the flag $f_i$ is set to one if and only if the corresponding price $p_i$ was selected by one of the bidders, the seller is able to know the set $\mathcal{P}' \subseteq \mathcal{P}$ of the selected prices, yet, he still does not know which bidder selected which price. The seller then publishes the winning price (bid). The bidder of the highest price announces himself or identifies himself to the seller and proves his case by opening the secret code corresponding to the highest price in the set $\mathcal{P}'$.

## 4. OVERVIEW OF THE BASIC TOOLS
In this section we introduce a brief overview of the underlying primitives used in our auction protocol.

### 4.1. Oblivious transfer
Rabin [26] proposed the concept of oblivious transfer (OT) in the cryptographic scenario. In this case The sender (seller) has only one secret bit $m$ and would like to have the receiver (buyer) to get it with probability 1/2, on the other hand, the receiver does not want the sender to know whether it gets $m$ or not. For $OT_2^1$, the sender has two secrets $m_1$ and $m_2$, the receiver will get one of them at the receiver's choice. The receiver does not want the sender to know which bit he chooses and the receiver must not know any information other than what he has chosen. $OT_k^1$ is a natural extension of the $OT_2^1$ to the case of $k$ secrets. However, constructing $OT_k^1$ from $OT_2^1$ is not a trivial problem. $OT_k^1$ is also known as "All or nothing disclosure of secrets (ANDOS)" [27, 28, 29, 30]. Oblivious transfers is a fundamental primitive in many cryptographic applications and secure distributed computations and has many applications such as private information retrieval (PIR), fair electronic contract signing, oblivious secure computation, etc. [31, 32]. In our auction protocol we will employ an efficient $OT_k^1$ oblivious transfer of strings [e.g. 2].

### 4.2. The Knapsack problem
The knapsack problem is a mathematically attractive proposition for cryptography. Merkle and Hellman public-key asymmetric cryptosystem [1] was based on the trapdoor knapsack problem. Assume a key $K = (k_1,...,k_\ell)$ where the $k_i$'s are integers and $\ell$ is the plaintext bit length. Let $X = (x_1,...,x_\ell)$ be the plaintext where, $x_i \in \{0,1\} \ \forall i = (1,...,\ell)$. Then the knapsack cryptosystem encrypts the plaintext $X$ according to the formula: $Y = K \bullet X = \sum_{i=1}^{\ell} k_i x_i$. The calculation of $Y$ from $X$ and $K$ is simple, while the recovery of $X$ from $Y$ and $K$ involves solving a knapsack problem and is generally difficult when $K$ is randomly chosen. If the key $K$ is chosen at random but also is chosen such that each element of $K$ is larger than the sum of the preceding elements, the corresponding knapsack problem becomes very simple. That is, if $k_i > \sum_{j=1}^{i-1} k_j \ \forall i = (1,...,\ell)$, the ciphertext can be generated as $y_\ell = \sum_{j=1}^{\ell} k_j x_j$. The plaintext $X$ can be recovered from the key $K$ and the ciphertext $y_\ell$ according to the following procedure: If $y_\ell < k_\ell$, then set $x_\ell = 0$ and $y_{\ell-1} = y_\ell$. If $y_\ell > k_\ell$, then set $x_\ell = 1$ and $y_{\ell-1} = y_\ell - k_\ell$. Using the computed value of $y_{\ell-1}$, the values $x_{\ell-1}$ and $y_{\ell-2}$ can be found in a similar fashion. The process continues until the whole $X$ is recovered. Our auction protocol will rely mainly on the knapsack problem.

### 4.3. Proof of equality of discrete logarithms
Given two large primes $p$ and $q$ such that $q \mid p-1$, $G_q$ is the unique multiplicative subgroup of $Z_p$ with order $q$. $g_1, g_2 \in G_q$. Alice and Bob knows $v, w, g_1, g_2$ but only Alice knows $x$ where $v = g_1^x$ and $w = g_2^x$. The proof of equality of the exponents is as follows:
- Alice chooses $z$ at random and sends $A = g_1^z$ and $B = g_2^z$ to Bob.
- Bob chooses a challenge $c$ at random and sends it to Alice.
- Alice sends $r = (z + cx) \bmod q$ to Bob.

Bob checks that $g_1^r = Av^c$ and $g_2^r = Bv^c$.

### 5. OUR KNAPSACK AUCTION PROTOCOL FOR HONEST PARTICIPANTS
In this section we give a concrete description of the proposed auction protocol when the participants (the seller and the bidders) incorporated in the protocol are honest but curious. The term 'honest' means that all the participants execute the steps of the protocol correctly and honestly. The term 'curious" means that the participants are willing to view any secret information leaked during the execution of the protocol but they do not deviate from the correct execution. This is equivalent to the eavesdropping adversary model. An eavesdropping adversary watches and learns all the information transferred to and from the corrupted participant but does not prevent him from contributing in the protocol correctly. *In the described protocol, we will assume that each bidder makes a bid different from any other bidder, that is, all bids are distinct. Soon, and after describing the tie-free auction protocol, we will show how to detect a tie between two or more bidders.* The tie-free auction protocol is given in the following subsection.

### 5.1. The detailed description of the protocol
*Initialization by the seller*:
- A good $G$ and its set of prices, $\mathcal{P} = \{p_1, ..., p_k\}$ where $p_i > p_j \; \forall i > j$ (ascending order).
- The seller defines a set of $k$ super-increasing random integers, $C = \{c_1, ..., c_k\}$ where $c_i > \sum_{j=1}^{i-1} c_j \; \forall i = (1, ..., k)$. He assigns $c_i$ to the price value $p_i$.
- The seller selects a set of $n$ random and large randomizing integers $\mathcal{R} = \{r_1, ..., r_n\}$ such that $\sum_{i=1}^{n} r_i = 0 \bmod q \; \wedge \; r_i \in Z_q^* \forall i = (1, ..., n)$ where $q$ is a large prime and $q > \sum_{i=1}^{k} c_i$.
- The seller defines a vector of $k$ flags, $\mathcal{F} = \{f_1, ..., f_k\}$ where $f_i \in \{0,1\}$ and initially, $f_i = 0 \forall i = (1, ..., k)$.

*Choices made by the bidders (distinct bids)*:
- Each bidder, $B_j \; \forall j = (1, ..., n)$ secretly chooses a price index $(i_j)$ corresponding to his selected price, $p_{i_j}$.

*Oblivious transfer of the chosen price secret code*:
- Each bidder $B_j$ interacts with the seller in an $OT_k^1$ oblivious transfer of strings to receive the secret randomized code, $c'_{i_j} = c_{i_j} + r_j \bmod q$ corresponding to his chosen price among the set of codes, $C_j = \{c_1 + r_j \bmod q, ..., c_k + r_j \bmod q\}$.

*Additive sharing of the secret codes*:
- Each bidder $B_j$ splits his secret code, $c'_{i_j}$ into $n$ random values, $d_{j,v}$ such that, $c'_{i_j} = \sum_{v=1}^{n} d_{j,v}$. This is spoken off as split knowledge.
- Each bidder $B_j$ privately sends $d_{j,v}$ to bidder $B_v \; \forall v = (1, ..., n)$.
- Each Bidder $B_j$ sums what he has from the other bidders to compute the additive share, $\sigma_j = \sum_{v=1}^{n} d_{v,j}$ and secretly sends $\sigma_j$ to the seller.
- The seller collects the additive shares from the bidders to compute $\sigma_k$.

*Solving the knapsack problem by the seller*:

- The seller computes the knapsack value, $\sigma_k = \sum_{i=1}^{n} \sigma_i \bmod q$.
- If $\sigma_k < c_k$, then set $f_k = 0$ and $\sigma_{k-1} = \sigma_k$. If $\sigma_k > c_k$, then set $f_k = 1$ and $\sigma_{k-1} = \sigma_k - c_k$.
- Using the computed value of $\sigma_{k-1}$, the seller sets $f_{k-1}$ and computes $\sigma_{k-2}$ in a way similar to the previous step.
- The seller continues solving the knapsack problem until all the states of the flags in $\mathcal{F}$ are determined.

*Announcing the winning bidder*:
- The seller broadcasts the highest price indicated by the vector $\mathcal{F}$ which contains the flags corresponding to the prices $\mathcal{P}$ chosen by the bidders.
- The seller requests the winning bidder (holding the secret code corresponding to the highest price) to identify himself.
- The winning bidder proves his case to the seller by showing the secret code he holds.

Although we described the protocol as a 1st price auction, one may notice that the protocol is automatically a 2nd price auction since the winning bidder can pay the 2nd highest price indicated by the flags vector, $\mathcal{F}$.

**5.2. Simple numerical illustrative example**
We will give a simple example at a medium level of details to illustrate the basic operations and computations. We will not describe the lower level details. The parameters are chosen strictly for simplicity of the example.

Consider a seller offering a good $G$ for 8 bidding prices, $\mathcal{P} = \{10\$, 20\$, 30\$, 40\$, 50\$, 60\$, 70\$, 80\$\}$. He chooses 8 super-increasing secret random integers in $Z_q^*$, $C = \{3, 5, 10, 21, 40, 90, 180, 360\}$. Choose the prime $q$ as $(q = 751) > 709$. The seller chooses 4 random and independent integers in $Z_q^*$ and sets the vector $\mathcal{R} = \{700, 100, 200, 502\}$ satisfying, $700 + 100 + 200 + 502 \equiv 0 \bmod 751$.

There are 4 bidders $\mathcal{B} = \{B_1, B_2, B_3, B_4\}$ willing to bid for $G$, each bidder makes his bid. Assume that the following bids are made: $B_1 \leftarrow 30\$, B_2 \leftarrow 10\$, B_3 \leftarrow 40\$, B_4 \leftarrow 80\$$. The seller interact with each bidder through an $OT_8^1$-oblivious transfer of strings to securely transfer the randomized secret code corresponding to the bidder's selected price index. As a result, each bidder possesses a randomized secret code corresponding to his selected price, that is,

$B_1 \leftarrow 710 \equiv 700 + 10 \bmod 751$,
$B_2 \leftarrow 103 \equiv 3 + 100 \bmod 751$,
$B_3 \leftarrow 221 \equiv 200 + 21 \bmod 751$,
$B_4 \leftarrow 111 \equiv 360 + 502 \bmod 751$.

Each bidder splits his secret code into 4 random additive shares,

$B_1 : 710 = 100 + 400 + 200 + 10$,
$B_2 : 103 = 10 + 50 + 40 + 3$,
$B_3 : 221 = 150 + 50 + 19 + 2$,
$B_4 : 111 = 30 + 35 + 35 + 11$.

Each bidder receives a share from each other bidder and computes his additive share of $\sigma_8$ as:

$B_1 : \sigma_1 = 100 + 10 + 150 + 30 = 290$,
$B_2 : \sigma_2 = 400 + 50 + 50 + 35 = 535$,
$B_3 : \sigma_3 = 200 + 40 + 19 + 35 = 294$,
$B_4 : \sigma_4 = 10 + 3 + 2 + 11 = 26$.

Each bidder $B_j$ sends his additive share $\sigma_j$ to the seller. The seller computes the knapsack value, $\sigma_8 = (290 + 535 + 294 + 26) \mod 751 = 394$.

The seller starts solving the knapsack problem and sets the set of flags $\mathcal{F}$. Given $C = \{3, 5, 10, 21, 40, 90, 180, 360\}$ and $\sigma_8 = 394$. The seller proceeds as follows:

$\sigma_8 > c_8 \Rightarrow f_8 = 1, \sigma_7 = \sigma_8 - c_8 = 394 - 360 = 34;$
$\sigma_7 < (c_7, c_6, c_5) \Rightarrow f_7 = f_6 = f_5 = 0, \sigma_4 = \sigma_5 = \sigma_6 = \sigma_7 = 34;$
$\sigma_4 > c_4 \Rightarrow f_4 = 1, \sigma_3 = \sigma_4 - c_4 = 34 - 21 = 13;$
$\sigma_3 > c_3 \Rightarrow f_3 = 1, \sigma_2 = \sigma_3 - c_3 = 13 - 10 = 3;$
$\sigma_2 < c_2 \Rightarrow f_2 = 0, \sigma_1 = \sigma_2 = 50, f_1 = 1.$

Finally the seller sets $\mathcal{F}$ = {1,0,1,1,0,0,0,1} and announces the highest price as 80\$. The winning bidder $B_4$ identifies himself to the seller and proves his case by showing the secret code, 710 to the seller, the winning bidder $B_4$ can pay the second highest price indicated by the flags which is 40\$ when the auction is a 2$^{nd}$ price auction.

## 6. OUR KNAPSACK AUCTION PROTOCOL FOR MALICIOUS PARTICIPANTS

Since the participants of the auction protocol are malicious, the first thing that comes to mind is that each participant must be committed to the values he selects, possesses or computes during the execution of the auction protocol in order to detect any attempt to manipulate or tamper with these values. As in most multiparty computation protocols a one way function with a suitable homomorphic property is required. Feldman in [33] used a one way function based on the discrete log problem. The main objective of Feldman was to add verifiability property to the well know Shamir's secret sharing scheme [34]. Let $p$ and $q$ be two large prime numbers such that $p = \mu q + 1$ or in other words $q | (p-1)$ where $\mu$ is a small integer. Let $g$ be an element of $Z_p$ and of order $q$ such that for each $x_i$ there is a public value $y_i = g^{x_i} \mod p$. Let $a_0, \ldots, a_t$ be the coefficients of Shamir's $t$-degree polynomial $f$ where, $a_0 = x$ is the secret. The dealer broadcasts $g^{a_0} = g^x, g^{a_1}, \ldots, g^{a_t}$. Each player $P_i$ can check the validity of his share $x_i$ by checking that, $g^{x_i} = (g^{a_0})(g^{a_1})^i \ldots (g^{a_t})^{i^t} \mod p$. Similar verification can be done during the reconstruction of the secret key to verify the validity of the submitted shares.

### 6.1. The detailed description of the protocol

In our auction protocol, we need two generators: $g_s$ which is used to commit the seller to the selected secret codes, and $g_b$ which is used to commit the bidders to their selected random values. Part of the protocol requires that the values published by the bidders must be away from the view of the seller. This can be achieved by many ways, for example, the bidders agree on a common secret key for a symmetric encryption algorithm (e.g., DES) and use this secret key to hide the published quantities from the seller. The auction protocol is as follows:

*Initialization by the seller*:
- A good $G$ and its set of prices, $\mathcal{P} = \{p_1, \ldots, p_k\}$ where $p_i > p_j \; \forall i > j$ (ascending order).
- The seller defines a set of $k$ super-increasing random integers, $C = \{c_1, \ldots, c_k\}$ where $c_i > \sum_{j=1}^{i-1} c_j \; \forall i = (1, \ldots, k)$. He assigns the secret code $c_i$ to the price value $p_i$.
- The seller selects a set of $n$ random and large randomizing integers $\mathcal{R} = \{r_1, \ldots, r_n\}$ such that $\sum_{i=1}^{n} r_i = 0 \mod q \; \wedge \; r_i \in Z_q^* \forall i = (1, \ldots, n)$ where $q$ is a large prime and $q > \sum_{i=1}^{k} c_i$.
- The seller defines a vector of $k$ flags, $\mathcal{F} = \{f_1, \ldots, f_k\}$ where $f_i \in \{0,1\}$ and initially, $f_i = 0 \forall i = (1, \ldots, k)$.
- The seller publishes $\eta_i = g_s^{c_i} \; \forall i = (1, \ldots, k)$ and $\xi_i = g_s^{r_i} \; \forall i = (1, \ldots, n)$.

*Choices made by the bidders*:
- Each bidder, $B_j \; \forall j = (1,...,n)$ secretly chooses a price index $(i_j)$ corresponding to his selected price $p_{i_j}$.

*Oblivious transfer and verification of the chosen price's secret code*:
- Each bidder $B_j$ interacts with the seller in an $OT_k^1$ oblivious transfer of strings to receive the secret code $c'_{i_j}$ corresponding to his chosen price.
- Each bidder $B_j$ verifies that the secret code he received from the seller is a valid code and corresponds to his index value by checking that, $\eta_i \xi_j = g_s^{c'_{i_j}}$.
- If the equality in the previous step does not hold, the bidder $B_j$ broadcasts a rejection, $REJ_j$. Else, he broadcasts an acceptance, $ACC_j$ and he cannot – later –repudiate the correct reception of the secret code.

*Committing the bidders to their chosen secret codes:*
- Away from the view of the seller, each bidder $B_j$ publishes a commitment to his secret code as $\zeta_j = g_b^{c'_{i_j}}$.

*Proof* 1: In the previous step, it is possible that a bidder lies and publishes a commitment unrelated to his secret code. Hence, it is required to prove that each published value $\zeta_j \forall j = (1,...,n)$ is valid (i.e. the exponent of $g_b$ is a valid secret code). The seller helps in this proof as follows:
- The seller computes the quantities $g_b^{c_1},...,g_b^{c_k}$ and publishes them in a random order. Also he publishes $g_b^{r_j} \; \forall j = (1,...,n)$ so that any bidder is able to compute $g_b^{c'_{i_j}}$ for any $i, j$.
- The bidders verify that each commitment value $\zeta_j \forall j = (1,...,n)$ equal to one of the quantities published by the seller in the previous step.

*Verifiable Additive sharing of the secret codes*:
　　In this part of the auction protocol, any quantities published by the bidders are away from the view of the seller unless otherwise stated. The protocol proceeds as follows:
- Each bidder $B_j$ splits his secret code, $c'_{i_j}$ into $n$ random values, $d_{j,v}$ such that, $c'_{i_j} = \sum_{v=1}^{n} d_{j,v}$. He also publishes the commitments $g_b^{d_{j,v}} \; \forall v(1,...,n)$.
- Each bidder $B_j$ privately sends $d_{j,v}$ to bidder $B_v \; \forall v = (1,...,n)$.
- Each bidder $B_j$ verifies the validity of what he receives from every other bidder $B_v$ by checking the commitments. $B_j$ broadcasts $ACC_j^{(v)}$ if he accepts the share, otherwise, he broadcasts $REJ_j^{(v)}$.
- After bad bidders are disqualified, each bidder $B_j$ sums what he has to compute his additive share $\sigma_j$.
- Each bidder $B_j$ publishes $g_b^{\sigma_j}$ (the seller views these quantities). It is obvious that the bidders can verify the validity of every published value, $g_b^{\sigma_j}$.
- Each bidder $B_j$ secretly delivers $\sigma_j$ to the seller.

*Verifying the knapsack quantity and solving the knapsack problem by the seller*:
- The seller checks the validity of the received additive shares $\sigma_1,...,\sigma_n$ by checking that $g_b^{\sigma_j}$'s are valid and matches the published commitments by the bidders.

- The seller safely computes the knapsack quantity, $\sigma_k = \sum_{i=1}^{n} \sigma_i \bmod q$.
- The seller starts solving the knapsack problem and sets the vector $\mathcal{F}$.
- The seller announces the highest bid.

*Announcing the winning bidder*:
- The seller requests the winning bidder (holding the secret code corresponding to the highest price) to identify himself.
- The winning bidder proves his case to the seller by showing the secret code he holds.

*Proof* 2: It is possible that the seller lies about the set of flags he broadcasted especially when the auction is a 2$^{nd}$ price payment, since he can lie about the 2$^{nd}$ highest price by increasing its amount and setting a wrong flag. It is required a proof that all the broadcasted flags are correct. To prove this, notice that: $\prod_{i=1}^{k} g_s^{c_i f_i} = g_s^{\sigma_k}$. Also the participants know $\prod_{i=1}^{n} g_b^{\sigma_i} = g_b^{\sigma_k}$. The seller is asked to prove to the winning bidder that the exponent of $A = \prod_{i=1}^{k} g_s^{c_i f_i}$ is equal to the exponent of $B = \prod_{i=1}^{n} g_b^{\sigma_i}$. The proof is as follows:

- The seller chooses a random value $z \in Z_q^*$ and privately sends $g_s^z, g_b^z$ to the winning bidder
- The winning bidder chooses a challenge $\omega \in Z_q^*$ at random and sends it to the seller.
- The seller sends $r = (z + \omega \sigma_k) \bmod q$ to the winning bidder.
- The winning bidder checks that, $g_s^r = g_s^z A^\omega$ and $g_b^r = g_b^z B^\omega$.

### 6.2. Tie detection and breaking
We described the auction protocol assuming that the bidders make distinct bids. However, if two or more bidders selected the same price $p_i$, they will be assigned the same value, $c_i$ and consequently, at the end of the protocol the seller will not be able to solve the knapsack problem correctly. Therefore, the bidders must be able to detect a tie before computing the additive shares of the knapsack value $\sigma_k$. Recall that each bidder $B_j$ has published the commitments, $g_b^{c'_{i_j}}$. If the seller publishes the values $g_b^{-r_j} \ \forall j = (1,...,n)$ then the bidders are able to compute $g_b^{c_{i_j}} = g_b^{c'_{i_j}} g_b^{-r_j} \ \forall j = (1,...,n)$. Hence, the bidders are able to detect a tie and solve it.

### 6.3. Notes on halted or disqualified bidders
It is possible that one or more bidders are halted, disconnected or disqualified due to the detection of malicious behavior. In this case we have three situations:

- A bidder is halted before the oblivious transfer of the secret code.
- A bidder is halted after transferring the secret code but before the additive secret sharing.
- A bidder is halted after the sharing of the secret codes.

In the first situation, the seller simply discards the random value in $\mathcal{R}$ corresponding to the halted bidder(s) and continues the execution of the protocol with the remaining bidders. In the second situation, the bids made by the halted bidder(s) are discarded. In the third situation, the bidders must re-share the secret code they have among the remaining bidders.
In our protocol we employed a trivial Joint additive secret sharing which represents an $(n-1, n)$-threshold secret sharing where the threshold is $t = n - 1$. Shamir's $(t, n)$-secret sharing scheme of [34] can be employed in order to avoid the re-sharing of the secret codes when a number no more than $n - (t + 1)$ bidders are halted. However, the protocol in this case cannot withstand the collusion of more than *t* malicious bidders, since it is possible that a bidder has many agents that work for him.

## 6. AN OPEN PROBLEM
As a way to improve privacy to the maximum, it would be nice if it is possible to let the seller solves the knapsack problem only for the highest price in the case of a $1^{st}$ price auction or for the $1^{st}$ and $2^{nd}$ highest prices in the case of a $2^{nd}$ price auction and hides all the rest of the bids from the seller and at the same time enables the detection of a lying seller. Our protocol hides the correspondence of the bids, since the seller is not able to know which bid belongs to which bidder.

## 7. CONCLUSIONS
In this paper we introduced a novel technique for electronic sealed-bid auction by employing the well known mathematical proposition, the knapsack problem to enable the seller to solve for the bids made by the bidders without having any information about which bid belongs to which bidder, therefore, the protocol is private and anonymous, the identity and the corresponding bids of the bidders – except the winning bidder – are kept unknown. The protocol is simple and of efficient execution time and data transfer complexity over previous protocols. The protocol does not involve any trusted parties or auctioneers. Only the seller and the set of bidders are able to come to an agreement on the selling price. We have shown the complete description of the protocol in the case of honest but curious participants, also, we have shown the protocol to detect malicious behavior of the participants. Our protocol is a $1^{st}$ price and a $2^{nd}$ price auction since the winning bidder can simply pay the $2^{nd}$ highest price indicated by the corresponding flag. We have shown also how a tie can be detected by the bidders.

## 8. REFERENCES
[1]. R. C. Merkle and M. E. Hellman, Hiding Information and Signatures in Trapdoor Knapsacks, *IEEE Trans. Inform. Theory*, vol. IT-24, no. 5, pp 525-530, 1798.
[2]. W. Tzeng, Efficient 1-out-of-n Oblivious Transfer Schemes, *In Proc. of PKC2002, LNCS2274*, pp 159-171, *Springer-Verlag*, 2002.
[3]. M. Franklin and M. Reiter, The design and implementation of a secure auction service, *IEEE Transactions on Software Engineering*, vol. 5(22), pp 302-312, 1996.
[4]. Y. Watanabe and H. Imai, Reducing the round complexity of a sealed-bid auction protocol with an off-line TTP. *In Proceedings of the 7th ACM conference on Computer and communications security*, pp 80-86, *ACM Press*, 2000.
[5]. K. Sakurai and S. Miyazaki, An anonymous electronic bidding protocol based on a new convertible group signature scheme, *Proc. 5th Australasian Conference on Information Security and Privacy (ACISP2000), LNCS1841*, pp 385-399, Springer-Verlag, 2000.
[6]. K. Sakurai and S. Miyazaki, A bulletin-board based digital auction scheme with bidding down strategy towards anonymous electronic bidding without anonymous channels nor trusted centers, *In Proc. International Workshop on Cryptographic Techniques and E-Commerce*, pp 180-187, 1999.
[7]. M. Jakobsson and A. Juels, Mix and match: Secure function evaluation via ciphertexts, *In Proceedings of Asiacrypt2000*, pp 162-177, 2000.
[8]. M. Kudo. Secure electronic sealed-bid auction protocol with public key cryptography, *IEICE Trans. Fundamentals*, vol. E81-A(1), pp20-27, 1998.
[9]. K. Viswanathan, C. Boyd, and E. Dawson, A three phased schema for sealed-bid auction system design. *In Australasian Conference for Information Security and Privacy, ACISP2000*.
[10]. F. Brandt, Cryptographic protocols for secure second-price auctions, In M. Klusch and F. Zambonelli, editors, *Cooperative Information Agents* V, vol. 2182 of *Lecture Notes in Artificial Intelligence*, pp. 154-165, *Berlin et al.*, 2001.
[11]. H. Kikuchi, (M+1)st-Price Auction Protocol, *In proceedings of Financial Cryptography* (*FC2001*), 2001.
[12]. H. Kikuchi, S. Hotta, K. Abe, and S. Nakanishi, Resolving winner and winning bid without revealing privacy of bids, *In Proceedings of the International Workshop on Next Generation Internet* (*NGITA*), pp 307-312, 2000.
[13]. H.Kikuchi, M.Harkavy and J.D.Tygar, Multiround Anonymous Auction Protocols, *In Proceedings of 1st IEEE Workshop on Dependable and Real-Time ECommerce Systems*, pp.62-69, 1998.
[14]. M.Harkavy, J.Tygar and H.Kikuchi, Electronic Auctions with Private Bids, *In Proceedings of 3rd USENIX Workshop on Electronic Commerce*, pp.61-74, 1998.
[15]. D. Song and J. Millen. Secure auctions in a publish/subscribe system, *Available at* http://www.csl.sri.com/users/millen/, 2000.
[16]. Olivier Baudron and Jacques Stern, Non-interactive Private Auctions. *In Paul Syverson, editor, Financial Cryptography - Fifth International Conference, Lecture Notes in Computer Science, Grand Cayman, BWI*, pp 19-22, Feb. 2001. *Springer-Verlag*.

[17]. C. Cachin, Efficient Private Bidding and Auctions with an Oblivious Third Party, *In Proceedings of the 6th ACM Conference on Computer and Communications Security*, pp 120-127, 1999.

[18]. H. Lipmaa, N. Asokan, and V. Niemi, Secure Vickrey Auctions without Threshold Trust, *In Financial Cryptography 2002, Lecture Notes in Computer Science, Southhampton Beach, Bermuda, 11-14 March 2002, Springer-Verlag.*

[19]. M. Naor, B. Pinkas, and R. Sumner, Privacy Preserving Auctions and Mechanism Design, *In Proceedings of ACM Conference on Electronic Commerce,* pp 120-127, 1999.

[20]. M.Abe and K.Suzuki, M+1-st Price Auction Using Homomorphic Encryption, *Proc. of Public Key Cryptography 2002*, *LNCS 2274*, pp 115-124, 2002.

[21]. M. Yokoo and K. Suzuki, Secure multi-agent dynamic programming based on homomorphic encryption and its application to combinatorial auctions. *In Proceedings of the First International Joint Conference on Autonomous Agents and Multi-Agent Systems, 2002,* to appear.

[22]. F. Brandt, Fully private auctions in a constant number of rounds, *In Proceedings of the 7th Annual Conference on Financial Cryptography (FC), Lecture Notes in Computer Science*, *Springer*, 2003, to appear.

[23]. F. Brandt, Secure and private auctions without auctioneers, *Technical Report FKI-245-02, Institut fur Informatik, Technische Universitat Munchen*, 2002.

[24]. F. Brandt, A Verifiable, Bidder-Resolved Auction Protocol. *In Proceedings of the 5th International Workshop on Deception, Fraud and Trust in Agent Societies,* pp 18--25, 2002.

[25]. F. Brandt. *Cryptographic protocols for secure second-price auctions. In M. Klusch and F. Zambonelli, editors, Cooperative Information Agents V,* vol. 2182 *of Lecture Notes in Artificial Intelligence,* pp. 154-165, Berlin et al., 2001, Springer.

[26]. M. Rabin, "How to exchange secrets by oblivious transfer," Technical Report TR-81, Harvard Aiken Computation Laboratory, 1981.

[27]. Y. Gertner, Y. Ishai, E. Kushilevitz and T. Malkin, "Protecting data privacy in information retrieval schemes," in *Proc. of $30^{th}$ Stoc*, 1998.

[28]. Julien P. Stern. "A new and efficient all-or-nothing disclosure of secrets protocol," in ASIACRYPT'98, pages 357-371. Springer-Verlag, 1998.

[29]. E. Kushilevitz and R. Ostrovsky. "Single-database computationally private information retrieval," in *Proc. of $38^{th}$ FOCS*, pages 364-373, 1997.

[30]. C. Cachin, S. Micali and M. Stadler, "Computa-tionally private information retrieval with polyloga-rithmic communication," in Advances in Crypto-graphy, EUROCRYPT'99, 1999.

[31]. M. Ben-Or, S. Goldwasser, A. Wigderson, "Completeness thearems for non-cryptographic fault-tolerant distributed computation," in *Proc. of the $20^{th}$ ACM symposium on the theory of computing*, pp. 1-10, 1988.

[32]. B. Chor, O. Goldreich, E. Kushilevitz, M. Susdan, "Private information retrieval," *Jornal of the ACM 45(6)*, pp. 965-982, 1998.

[33]. P. Feldman, A Practical Scheme for Non Interactive Verifiable Secret Sharing, *In Proc. of the 28th IEEE Symposium on the Foundations of Computer Science*, pp. 427-437, 1987.

[34]. A. Shamir, How to Share a Secret, *Communications of the ACM,* 22(11)*,* pp. 612-613, *ACM,* November 1979.

[35]. K Sakurai, S Miyazaki, A bulletin-board based digital auction scheme with bidding down strategy, *International Workshop on Cryptographic Techniques and E-Commerce*, 2009